\documentclass[vecphys]{svmult}

\usepackage{makeidx}         % allows index generation
\usepackage{graphicx}        % standard LaTeX graphics tool
\usepackage{multicol}        % used for the two-column index
\usepackage{cite}            % adjusts the "syntax" of the refs in the
\usepackage[bottom]{footmisc}% places footnotes at page bottom

\usepackage{color}

\newcommand{\La}{\line (1,0  ){12}}
\newcommand{\Lb}{\line (3,5 ){6}}

\newcommand{\Ld}{\line (-1,0){12}}
\newcommand{\Le}{\line (-3,-5){6}}

\newcommand{\C} {\circle*{4}}

\newcommand{\pA}{\put(-6,-10)}
\newcommand{\pB}{\put(6,-10)}
\newcommand{\pC}{\put(12,0)}

\newcommand{\pZ}{\put(0,0)}

\newcommand{\rhomb}{
  \pA{\C}\pB{\C}\pZ{\C}\pC{\C}
 }

\newcommand{\rhombH}{
  \begin{picture}(22,10)(-8,-6)
    \rhomb
    \pA{\La}\pC{\Ld}
  \end{picture}
}
\newcommand{\rhombV}{
  \begin{picture}(22,10)(-8,-6)
    \rhomb
    \pB{\Lb}\pZ{\Le}
  \end{picture}
}
\makeindex             % used for the subject index
\begin{document}

\title{Strong-coupling expansion and effective hamiltonians}

\author{Fr\'ed\'eric Mila$^1$ \and Kai Phillip Schmidt$^2$}

\institute{$^1$Institute of Theoretical Physics, Ecole Polytechnique
F\'ed\'erale de Lausanne, CH-1015 Lausanne, Switzerland;
\texttt{frederic.mila@epfl.ch}\\
$^2$Lehrstuhl f\"ur Theoretische Physik I, TU Dortmund, D-44221
Dortmund, Germany; \texttt{schmidt@fkt.physik.uni-dortmund.de}}

\maketitle

When looking for analytical approaches to treat frustrated quantum magnets,
it is often very useful to start from a limit where the ground state is
highly degenerate. This chapter discusses several ways of deriving { effective Hamiltonians} around such limits, starting from standard { degenerate
perturbation theory} and proceeding to modern approaches more appropriate
for the derivation of high-order effective Hamiltonians, such as the
perturbative continuous unitary transformations or contractor
renormalization. In the course of this exposition, a number of examples
taken from the recent literature are discussed, including frustrated
ladders and other dimer-based Heisenberg models in a field, as well as
the mapping between frustrated Ising models in a transverse field and
quantum dimer models.

\section{Introduction}

As emphasized several times throughout this book, frustrated magnets often
have a highly degenerate ground state in the classical limit. This is
sometimes even taken as a definition of highly frustrated magnets. This
degeneracy makes the semi-classical expansion in $1/S$ effectively
uncontrolled (if it does not already fail simply because of divergent
quantum fluctuations), because usually one cannot perform and thus compare
the expansions around all classical ground states. An infinite degeneracy
is also often present in other limiting cases such as decoupled local units
(such as triangles in the $S$ = 1/2 trimerized kagome lattice) or the Ising
limit (for systems such as the antiferromagnetic Heisenberg model on the
triangular or the kagome lattice). In such limits, which preserve the
quantum nature of the problem, this degeneracy is not the end of the story,
but rather the starting point of a systematic expansion, namely degenerate
perturbation theory, which leads to an effective Hamiltonian. In the context
of strongly correlated systems, this type of method usually goes by the
name '{ strong-coupling expansion},' because the starting point of the
perturbative expansion is a Hamiltonian where only the interaction terms
are kept, the kinetic terms being treated as the perturbation.

There are several ways to perform this expansion, or more generally to
derive an effective Hamiltonian. There are in fact two types of effective
Hamiltonian: those which act only in the degenerate subspace of a
non-perturbed Hamiltonian, and those which act in the full Hilbert space
but, through a canonical transformation, are rewritten as a series in terms
of the ratio of two parameters. While the first type can always be written
down explicitly, the second type can be derived in a simple way only provided
that one can find a suitable generator. In the following, we will discuss
both types of effective Hamiltonian, starting with the expansion in the
degenerate subspace because this is more standard.

The derivation of an effective Hamiltonian is extremely useful in isolating
the relevant degrees of freedom. However, the problem is usually not solved
once the effective Hamiltonian has been derived. Indeed, the new Hamiltonian
often poses a problem as difficult as the original one. The primary advantage
of the effective Hamiltonian is that, because the relevant degrees of freedom
have been selected, simple approximations to the problem defined by the
effective Hamiltonian often give deep insight into the physics of the
problem. We will illustrate this point with several examples.

This chapter is organized as follows. In Sec.~II, we review briefly
the degenerate perturbation theory approach to effective Hamiltonians,
with a concise but self-contained discussion of the second-order result,
and a description of the form the expansion takes at higher orders. In
Sec.~III, we discuss three examples taken from the field of frustrated
magnetism where this approach has proven very useful: coupled dimers in a
magnetic field, the Ising model in a transverse field, and the trimerized,
spin-1/2 kagome antiferomagnet. In Sec.~IV, we review more
sophisticated approaches based on the same foundation: canonical
transformations, continuous unitary transformations (CUTs), and the
contractor renormalization group approach (CORE). Note that linked cluster
expansions are discussed in another chapter of this book (that by L\"auchli).
We conclude in Sec.~V with a discussion that includes a comparison of the
various approaches.

\section{Strong-coupling expansion}

Let us consider a system described by a Hamiltonian
$$
H = H_0 + V$$
acting in a Hilbert space $\cal H$ such that the ground state of $H_0$
is degenerate. We denote by ${\cal H}_0$ the Hilbert space of the
ground-state manifold. The goal is to find an effective Hamiltonian
$H_{\rm eff}$ acting in ${\cal H}_0$ such that
$$ H_{\rm eff} \vert \phi \rangle = E \vert \phi \rangle \Rightarrow H
\vert \psi \rangle = E \vert \psi \rangle,\ \ \
\vert \phi \rangle \in {\cal H}_0,\ \ \ \vert \psi \rangle \in {\cal H}.
$$

\subsection{{ Second-order perturbation theory}}

Up to second order, the relation to be derived below is a standard result of quantum mechanics
\cite{landau}. Denoting by $E_0$ the ground-state energy of $H_0$
and by $E_m$ the other eigenenergies, $H_0\vert m \rangle = E_m \vert m
\rangle$ for $E_m \ne E_0$. Thus for two vectors $\vert \phi \rangle,\,
\vert \phi' \rangle \in {\cal H}_0$, and up to second order in $V$,
$$
\langle \phi \vert H_{\rm eff} \vert \phi' \rangle = \langle \phi
\vert H_0 \vert \phi' \rangle + \langle \phi \vert V \vert \phi'
\rangle + \sum_{\vert m \rangle \notin {\cal H}_0} \frac{\langle
\phi \vert V \vert m \rangle \langle m \vert V \vert \phi' \rangle}{E_0-E_m}.
$$
This result can be reformulated as an operator identity \cite{fulde}. If
one denotes by $P$ the projector on ${\cal H}_0$, and defines $Q = 1 - P$,
then to second order in $V$
$$
PH_{\rm eff}P = P H_0 P + P V P + P V Q\frac{1}{E_0-QH_0Q}Q VP.
$$

\noindent
{\bf Proof:} suppose $H \vert \psi \rangle = E \vert \psi \rangle$. Because
$P + Q = 1$, this can be written as
$$
(P + Q) H (P + Q) \vert \psi \rangle = E \vert \psi \rangle.
$$
Projecting onto ${\cal H}_0$ and ${\cal H}-{\cal H}_0$ gives
$$
PHP \vert \psi \rangle + PHQ \vert \psi \rangle= E P \vert \psi
\rangle, \ \ \ \ (1)
$$
$$
QHP \vert \psi \rangle + QHQ \vert \psi \rangle= E Q \vert \psi \rangle,
\ \ \ \ (2)
$$
whence
$$
(2) \Rightarrow Q \vert \psi \rangle= (E-QHQ)^{-1}QHP\vert \psi \rangle,
$$
$$
(1) \Rightarrow PHP \vert \psi \rangle + PHQ\frac{1}{E-QHQ}Q HP \vert
\psi \rangle = E P \vert \psi \rangle.
$$
Expansion of $(E - QHQ)^{-1}$ using $(A-B)^{-1} = A^{-1} \sum_{n=0}^{\infty}
\left( BA^{-1}\right)^n$, with $A = E_0 - Q H_0 Q$ and $B = Q V Q - E + E_0$,
leads to
\begin{eqnarray}
PH_{\rm eff}P & = & P H_0 P + P V P + P V Q (E_0 - QH_0Q)^{-1}\nonumber \\
& & \times \sum_{n=0}^{\infty} \left( ( Q V Q- E + E_0)
(E_0 - QH_0Q)^{-1} \right)^n Q V P.
\end{eqnarray}
The expansion is then truncated at $n = 0$. QED.

\subsection{{High-order perturbation theory}}

This form of the expansion is not well suited to the derivation of
higher-order expansions, because beyond second order it contains explicitly
the exact eigenenergy $E$. An expansion only in terms of the unperturbed
eigenenergies can nevertheless be derived. The first systematic method for
this dates to the work of Kato\cite{kato}. Here we follow the formulation
of Takahashi \cite{takahashi}, in which the expansion takes the form
$$P H_{\rm eff} P = \Gamma^\dagger H \Gamma,$$
$$ \Gamma = \bar P P (P \bar P P)^{-1/2},$$
$$(P \bar P P)^{-1/2} = P + \sum_{n=1}^\infty \frac{(2n-1)!!}{(2n)!!}
[P(P-\bar P)P]^n,$$
$$\bar P = P - \sum_{n=1}^\infty \ \ \sum_{k_1+...+k_{n+1}=n, k_i\ge 0}
S^{k_1}VS^{k_2}V...VS^{k_{n+1}},$$
$$S^0 = - P,\ \ S^k = \left( \frac{Q}{E_0-QH_0Q} \right)^k. $$
The true eigenstates $\psi$ are related to the eigenstates $\phi$ of
$H_{\rm eff}$ by
$$\vert \phi \rangle = \Gamma \vert \psi \rangle,$$
and likewise the observables transform according to
$$O \rightarrow \Gamma^\dagger O \Gamma.$$
Thus the $n$th order term of $H_{\rm eff}$ has the form
$$H_{\rm eff}^{(n)} = \sum_{k_1+ \dots +k_{n-1} = n-1, k_i\ge 0}
f(k_1,k_2,\dots,k_{n-1}) V S^{k_1} V S^{k_2} V...S^{k_{n-1}}V,$$
where the coefficients $f(k_1,\dots,k_{n-1})$ are deduced by appropriate
bookkeeping from the previous expansions. The number of terms in such a
strong-coupling expansion grows exponentially with $n$. In practice,
beyond the fourth order it can generally be carried out only with the
help of a computer. An alternative formulation based on continuous
unitary transformations, which is simpler when applicable, will be
discussed in the next section. In the remainder of this section, we
discuss a number of selected examples where low-order degenerate
perturbation theory provides considerable additional insight into
the problem.

\subsection{Examples}

In the field of quantum magnetism, the best known example is the derivation
of the Heisenberg model starting from the half-filled {Hubbard model}. The
Hubbard model is defined by
$$
H = V + H_0 = -t \sum_{\langle i,j \rangle, \sigma} (c^{\dagger}_{i\sigma}
c_{j\sigma} + h.c.) + U \sum_i n_{i\uparrow} n_{i\downarrow}.
$$
At half-filling, the ground state of the interaction term $H_0$ is $2^N$-fold
degenerate, where $N$ is the number of sites. Treating the kinetic term $V$
as a perturbation leads at second order (up to a constant) to the effective
Hamiltonian
$$H_{\rm eff} = J \sum_{\langle i,j \rangle} \vec S_i \cdot \vec S_j,$$
with $J = 4t^2/U$ \cite{anderson}. As noted in the introduction, this is
a case where the effective Hamiltonian is itself very difficult to solve,
which is indeed true for the {Heisenberg model} on most lattices. As we
shall see in the following, it is often useful to go one step further,
and to derive a further effective Hamiltonian starting from one
physically relevant limit.

\subsubsection{Example 1: frustrated spin-1/2 ladder in a magnetic field}

\begin{figure}[t]
\centering
\includegraphics*[width=.9\textwidth]{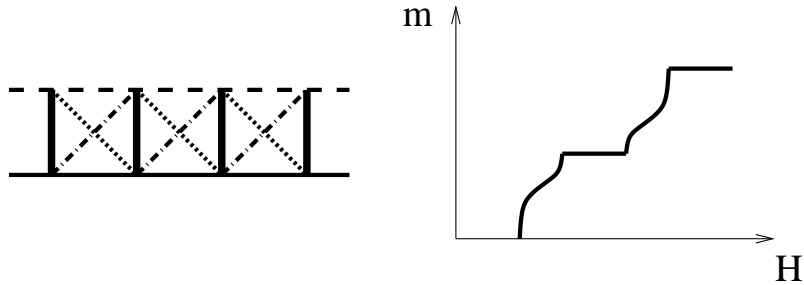}
\caption[]{Left: schematic representation of the spin ladder of
Eq.~\ref{eq:ladder}. The couplings entering Eq.~\ref{eq:ladder} are
denoted by the different line types: thick solid ($J_\perp$), thin solid
($J_{11}$), dashed ($J_{22}$), dot-dashed ($J_{12}$) and dotted ($J_{21}$).
Right: magnetization curve of a frustrated spin ladder with a
1/2-magnetization plateau.}
\label{fig:1}
\end{figure}

We consider the Heisenberg model for a frustrated spin-1/2 {ladder} in a
magnetic field defined by the Hamiltonian
\begin{equation}
H = \sum_n J_\perp \ \vec S_{1n}\cdot \vec S_{2n} - B \sum_n (S_{1n}^z +
S_{2n}^z)
+\sum_n \sum_{i,j=1,2} J_{ij} \ \vec S_{in}\cdot \vec S_{jn+1}.
\label{eq:ladder}
\end{equation}
In the spin operators $\vec S_{in}$, the index $i$ refers to the leg
and the index $n$ to the rung (Fig.~\ref{fig:1}). The goal is to derive
an effective Hamiltonian that describes the magnetization process in the
limit $J_\perp \gg J_{ij}$ \cite{totsuka,mila}.

The starting point of the perturbative expansion is the Hamiltonian of
isolated {dimers} at the critical field $B_c = J_\perp$ where, for one dimer,
the triplet polarized in the field direction crosses the singlet,
\begin{equation}
H_0 = \sum_n J_\perp \vec S_{1n} \cdot \vec S_{2n} - B_c \sum_ n (S_{1n}^z
 + S_{2n}^z).
\end{equation}
The ground state of this Hamiltonian is $2^N$-fold degenerate, where $N$
is the number of ladder rungs, and the full Hamiltonian can be treated
within degenerate perturbation theory, the perturbation being given by
\begin{equation}
V = \sum_{\langle nm \rangle} \sum_{i,j=1,2} J_{ij} \, \vec S_{in}\cdot
\vec S_{jm} - (B - B_c) \sum_n (S_{1n}^z + S_{2n}^z).
\end{equation}
The sum over $\langle nm \rangle$ refers to the nearest-neighbor rungs.
A perturbative expansion can be performed under the condition that the
matrix elements of $V$ are small compared to the excited states of $H_0$,
i.e. as long as $J_{ij}, (B - B_c) \ll J_\perp$. The condition on $B - B_c$
might suggest that such a calculation cannot give access to the full
{magnetization curve}, but in fact the magnetization is rigorously equal
to zero or to the full saturation value outside a window whose width is
of order $J_{ij}$. Thus this type of perturbation theory can indeed give
the full magnetization curve.

To write down the effective Hamiltonian, one needs a description of the
Hilbert space. Because there are two states per rung, one simple choice
is to introduce the Pauli matrices $\vec \sigma_n$ at each rung such that
the singlet corresponds to $\vert \sigma_n^z = - 1/2\rangle$ and the
triplet polarized along the field to $\vert \sigma_n^z = 1/2 \rangle$.
In this basis, and up to first order in $V$, the effective Hamiltonian
is given by
\begin{equation}
H_{\rm eff} = J^{xy} \sum_n (\sigma^x_n \sigma^x_{n+1} + \sigma^y_n
\sigma^y_{n+1}) + J^z \sum_n \sigma^z_n \sigma^z_{n+1} - B_{\rm eff}
\sum_n(\sigma^z_n) + C,
\end{equation}
with
\begin{eqnarray}
J^{xy} & = & \frac{1}{2}(J_{11} + J_{22} - J_{12} - J_{21}),\\
J^{z} & = & \frac{1}{4}(J_{11} + J_{22} + J_{12} + J_{21}),\\
B_{\rm eff} & = & B - B_c - \frac{1}{4}(J_{11} + J_{22} + J_{12} + J_{21}),\\
C & = & \frac{1}{8}(J_{11} + J_{22} + J_{12} + J_{21}) - (B - B_c).
\end{eqnarray}
The effective Hamiltonian is identical to that of an {XXZ chain} in a field,
which is a major step forward with respect to the original problem: this
model has been investigated at length using the Bethe ansatz \cite{bethe}
and by field-theory methods \cite{luther,haldane}, and much is known about
its low-energy properties. In particular, there is a quantum phase
transition in zero field at $J^z = J^{xy}$ between a gapless phase at
small $J^z$ and a gapped phase at large $J^z$.

To understand the physics of this phase transition for the original problem,
it is expedient to perform a {Jordan-Wigner transformation} \cite{mattis} of
the effective model. In this language, the Hilbert space is that of
spinless fermions on a chain, and the elementary operators are fermion
creation ($c_i^\dagger$), annihilation ($c_i$) and density ($n_i =
c^\dagger_i c_i$) operators at site $i$. An empty site corresponds to
a rung singlet, an occupied one to a rung triplet, and the effective
Hamiltonian becomes
\begin{equation}
H_{\rm eff} = -t \sum_i (c^\dagger_i c_{i+1} + c^\dagger_{i+1}c_i) +
v \sum_i n_i n_{i+1} - \mu \sum_n n_i,
\end{equation}
with $t = J^{xy}/2$, $v = J^z$, and $\mu = B_{\rm eff} + J^z$.
In this model, the gapped phase of the ladder is a half-filled insulating
phase of the fermionic chain, while the gapless one is a metallic phase
(a {Luttinger liquid} in this one-dimensional (1D) system). Thus the density
as a function of the chemical potential has a {plateau}, the width of which
is equal to the gap of the insulating phase. In the original model, this
implies that the magnetization can have a plateau for certain parameters,
a result which has been confirmed by density-matrix renormalization-group
(DMRG) calculations \cite{fouet}. The physics of the plateau state is
discussed elsewhere in this volume (the chapter by Takigawa and Mila).
For the purposes of the present chapter, we note only how powerful the
effective-Hamiltonian method can be: a very simple, first-order calculation
can basically solve the problem by mapping it onto another non-trivial but
well-understood one.

\subsubsection{Example 2: expansion around the Ising limit}

In the previous example, as in the case of the Hubbard model, the
unperturbed Hamiltonian is a sum of local terms, and the macroscopic
ground-state degeneracy is given simply from the ground-state degeneracy
of each term. This is not the only case where the effective-Hamiltonian
approach is useful. Another important example is the ground state of the
antiferromagnetic Ising model, which is degenerate on non-bipartite
lattices such as the triangular and kagome geometries. Again this can
be the starting point of a degenerate perturbation theory towards the
Heisenberg model on the same lattice if the transverse exchange is
treated as a perturbation. In the same spirit, degenerate perturbation
theory can be used to treat the effect of a transverse magnetic field
applied to a {frustrated Ising model}. In this section we concentrate on
the Ising case, which is of direct relevance to Quantum Dimer Models
(QDMs, discussed in the chapter by Moessner and Raman).

\begin{figure}[t]
\centering
\includegraphics*[width=.8\textwidth]{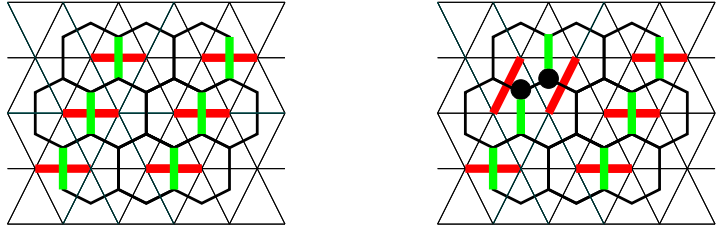}
\caption[]{Left: representation of the model of Eq.~\ref{eq:imtf}. The
thick vertical lines on the honeycomb lattice correspond to $M_{ij} = -1$,
all others to $M_{ij} = 1$. The thick horizontal lines on the triangular lattice
correspond to dimers in the ferromagnetic ground state. Right: dimer
covering obtained from the ferromagnetic ground state after flipping
the two spins shown as black dots.}
\label{fig:2}
\end{figure}

Consider the fully frustrated Ising model in a transverse field on the
honeycomb lattice, defined by the Hamiltonian
\begin{equation}
H = H_0 + V = - J \sum_{\langle i,j \rangle} M_{ij} \sigma^z_i \sigma^z_j -
\Gamma \sum_i \sigma^x_i.
\label{eq:imtf}
\end{equation}
In this expression, the parameters $M_{ij} = \pm 1$ are chosen in such a
way that their product around each hexagonal plaquette is equal to $-1$.
All models in this class are then equivalent up to a gauge transformation,
$\sigma_i^z \rightarrow \epsilon_i\sigma_i^z$, $M_{ij} \rightarrow
\epsilon_i \epsilon_j M_{ij}$, with $\epsilon_i = \pm 1$. One possible
choice is shown in Fig.~\ref{fig:2}. For this particular choice, the two
ferromagnetic states, with all spins either up or down, are ground states
of $H_0$. Indeed, they satisfy five bonds out of every six, and it is
clearly impossible to do better for a case with one antiferromagnetic
coupling out of six. From a ferromagnetic ground state, one may then
construct more ground states, and it is easy to verify that flipping the
spins at the ends of a satisfied bond that connects two unsatisfied bonds
leads to another ground state (right panel, Fig.~\ref{fig:2}).

The structure of the Hilbert space is best understood by considering the
{dual lattice}, which is the triangular lattice of sites at the centers of
the hexagonal plaquettes, and to draw a line between two neighboring
sites of the dual lattice if the bond (of the original lattice) which it
crosses is unsatisfied. If one imposes the constraint that each site be
connected to exactly one of its neighbors, this set of lines defines a
{dimer covering} of the triangular lattice, and there is a one-to-one
correspondence between these dimer coverings and the ground states of
$H_0$ which have a given spin orientation for each given site. Because
there are two possible spin orientations, the degenerate manifold of
$H_0$ is twice as large as the Hilbert space of dimer coverings, the
two configurations leading to the same dimer covering being related
by a global flip of all spins.

For simplicity, we focus on the problem defined by the Ising Hamiltonian
acting in the Hilbert space where two configurations related by a global
spin-flip are identified. Then the degenerate subspace can be described
by the set of dimer coverings, and the effective Hamiltonian obtained by
degenerate perturbation theory takes the form of a QDM. To deduce the form
of this QDM, we first note that the effective Hamiltonian vanishes to first
order in the transverse field. Indeed, flipping a single spin changes the
nature of the three bonds connected to it, which increases the number of
unsatisfied bonds by one or by three, depending on whether or not one of
the bonds connected to this site was unsatisfied. To second order, the
situation is still more complex unless the two sites are nearest neighbors,
in which case the process does not result in a state outside the
ground-state manifold provided the two sites are on a satisfied bond
connected to two unsatisfied ones. The resulting process leads to a dimer
flip around a square plaquette of the dual lattice, and the effective
Hamiltonian is a special case of the {Rokhsar-Kivelson model}, defined by
the Hamiltonian
\begin{eqnarray}
  \mathcal{H} = & - t & \sum_r \left(
  \left| \rhombV \right> \left< \rhombH \right| + {\rm H.c.}
  \right) \nonumber \\
  & + v & \sum_r \left(
  \left| \rhombV \right> \left< \rhombV \right| + \left| \rhombH \right>
\left< \rhombH \right|
  \right)
\label{eq:tQDM}
\end{eqnarray}
with $t = \Gamma^2/J$ and $v = 0$.

In this example, it is the effective model which is of direct interest in
the context of frustrated magnets (as discussed in the chapter by Moessner
and Raman). The connection with the Ising model turns out to be helpful in
identifying a possible phase transition, as first noted in
Ref.~\cite{chandra}, and led more recently to an analytical
description of the {fractional excitations} which exist in the
Resonating-Valence-Bond phase of the QDM on the square lattice
\cite{misguich}.

As mentioned at the beginning of this section, an expansion around the
Ising limit can also be performed for the XXZ version of the Heisenberg
model, using the ratio $J^{xy}/J^z$ as a small parameter. This approach
has been used for the 1/3-plateau of the spin-1/2 kagome antiferromagnet
\cite{cabra}, for the half-magnetization plateau phase of the spin-1/2
pyrochlore Heisenberg antiferromagnet \cite{bergman}, and for the
1/3-plateau phase of molecular analogs of the spin 1/2 kagome
antiferromagnet \cite{rousochatzakis}.

\subsubsection{Example 3: weakly {coupled triangles}}

Frustration naturally appears in antiferromagnets in which the exchange
paths create loops of odd length, the simplest of these loops being the
triangle. There are several types of lattice geometry which can be
considered as coupled triangles, and for which a perturbation theory
starting from non-interacting triangles has proven to be useful. The
effective model takes a special form due to the pecularity of the
ground-state manifold of a triangle, which is four-fold degenerate.
This is easily seen by rewriting the Hamiltonian in terms of the total
spin of the triangle, $\vec S_{\rm tot} = \vec S_1 + \vec S_2 + \vec S_3$,
to obtain
\begin{equation}
H = J(\vec S_1 \cdot \vec S_2 + \vec S_2 \cdot \vec S_3 + \vec S_3 \cdot
\vec S_1) = \frac{J}{2} \left[ (\vec S_{\rm tot})^2 - \frac{9}{4} \right],
\end{equation}
a result emerging because it is possible to construct two doublets, i.e.
four states, using three spin-1/2 entities. A convenient basis is provided
by the simultaneous eigenstates of the {scalar chirality}, $\vec{S_1}.
(\vec{S_2} \times \vec{S_3})$, and of the projection $S_{\rm tot}^z$ of
the total spin,
\begin{eqnarray}
|R, \sigma\rangle & = & (|-\sigma \sigma \sigma\rangle
+\omega |\sigma -\sigma \sigma\rangle + \omega^2
|\sigma \sigma -\sigma\rangle)/\sqrt{3}, \nonumber \\
|L, \sigma\rangle & = & (|-\sigma \sigma \sigma\rangle
+\omega^2 |\sigma -\sigma \sigma\rangle + \omega
|\sigma \sigma -\sigma \rangle)/\sqrt{3}, \nonumber
\end{eqnarray}
where $\omega = \exp(2i\pi/3)$, $\sigma = \pm 1/2$ refers to $S_{\rm tot}^z$,
and $L$ and $R$ represent left- and right-handed chirality.

In a system of weakly coupled triangles, treating the inter-triangle coupling
as a perturbation leads to an effective Hamiltonian which acts in a Hilbert
space of dimension $4^{N_t}$, where $N_t$ is the number of triangles. For a
given triangle $i$, it is convenient to introduce a spin-1/2 operator
$\vec{\sigma}_i$, acting on the total spin, and a Pauli-matrix vector
$\vec{\tau}_i$ acting in chirality space. To first order in the
inter-triangle couplings, provided these couplings are SU(2)-invariant,
the effective Hamiltonian then takes the general form
\begin{equation}
H_{\rm eff} = J_{\rm eff}{\sum_{i,j}}' \vec \sigma_i \cdot \vec \sigma_j
H_{ij}^\tau,
\end{equation}
where the sum is over the lattice of sites representing the arrangement of
the triangles, $J_{\rm eff}$ is linear in the inter-triangle couplings, and
the operator $H_{ij}^\tau$ depends on which sites of the triangles $i$ and
$j$ are coupled by the inter-triangle interaction.

In the case of a three-leg '{spin tube},' the effective model takes the
explicit form \cite{schulz}
\begin{equation}
H_{\rm eff}= \frac{J'}{3} {\sum_{\langle ij\rangle}}' \vec \sigma_i \cdot
\vec \sigma_j [1 + 2(\tau_i^x \tau_{i+1}^x + \tau_i^x \tau_{i+1}^x)],
\end{equation}
where $J'$ is the rung coupling of the real model. For the effective model,
field-theory arguments based on bosonization show that the spectrum must be
gapped in all sectors, a non-trivial prediction to be contrasted with the
gapless spectrum of the three-leg ladder \cite{schulz}.

Another example where this type of effective Hamiltonian has provided
additional valuable insight is on the {kagome lattice}. Considering the
trimerized version of this model, where the exchange constant within
up-pointing triangles is taken to be $J$ and in down-pointing triangles
$J'$, the effective Hamiltonian in the limit $J'\ll J$ can be expressed
as \cite{subrahmanyam,mila2}
\begin{equation}
H_{\rm eff}= \frac{J'}{9} {\sum_{\langle ij\rangle}}' \vec \sigma_i \cdot
\vec \sigma_j (1-4\vec e_{ij} \cdot \vec \tau_i)(1 - 4\vec e_{ij} \cdot
\vec \tau_j),
\end{equation}
where the vectors $\vec e_{ij}$ are taken from among $\vec e_1 = (1,0)$,
$\vec e_2 = (-1/2,-\sqrt{3}/2)$, and $\vec e_3 = (-1/2,\sqrt{3}/2)$, the
choice depending on the labelling of the spins inside each triangle
\cite{ferrero}. For this effective Hamiltonian, a mean-field decoupling
of spin and chirality leads to a highly degenerate ground state. Each
ground state can be associated with a dimer covering of the triangular
lattice of up- (or down-)pointing triangles, and the number of states
accordingly grows as $1.5351^{N_t}$ in terms of the number of triangles,
or as $1.1536^N$ in terms of the number of spins\cite{mila2}. This
compares well with the number of low-lying singlets observed in numerical
simulations of the spin-1/2 kagome antiferromagnet \cite{lecheminant}.

This effective Hamiltonian is also a very useful starting point to discuss
the physics in a {magnetic field}. Antiferromagnets composed of weakly coupled
triangles exhibit a plateau at magnetization 1/3, in which all triangles have
a total spin equal to 1/2 (in a $S = 1/2$ system) and oriented in the field
direction. Inside the plateau, $\vec \sigma_i \cdot \vec \sigma_j = 1/4$ is
a constant, and the effective Hamiltonian is a pure chirality model. This
plateau has been investigated for the kagome lattice in Ref.~\cite{honecker}.

Finally, similar ideas based on weakly coupled tetrahedra have provided
equally helpful insight into the properties of the spin-1/2 Heisenberg model
on the pyrochlore lattice and in related systems \cite{tsunetsugu,kotov}.

\section{Alternative approaches yielding effective Hamiltonians}

There are several ways of deriving {effective Hamiltonians} based on
techniques other than direct strong-coupling expansions. The aim of this
section is to provide a review of the physics underlying these approaches,
with appropriate references for further reading concerning their more
detailed implementations.

\subsection{Canonical transformation}

The {canonical transformation} of a Hamiltonian is defined by
$H \rightarrow \tilde{H} = e^{\eta}He^{-\eta}$, where $\eta$ is
antihermitian, so that $e^{\eta}$ is unitary. If $\vert \psi \rangle$
is an eigenstate of $H$, then $\vert \tilde{\psi} \rangle = e^{\eta}
\vert \psi \rangle$ is an eigenstate of $\tilde H$ with the same
eigenvalue. If the operators are transformed simultaneously according to
$A \rightarrow \tilde{A} = e^{\eta} A e^{-\eta}$, then $\langle\tilde{\psi}
\vert \tilde{A} \vert \tilde{\psi} \rangle = \langle \psi \vert A \vert
\psi \rangle$.

The foundation for using a canonical transformation to derive an effective
Hamiltonian is the identity
\begin{eqnarray}
e^{\eta} H e^{-\eta} & = & H + [\eta,H] +\frac{1}{2 !}[\eta,[\eta,H]] + \dots
\nonumber\\ & = & H + \sum_{n=1}^{\infty} \frac{1}{n!} [\eta,[\eta,\dots[
\eta,H] \dots ]],
\nonumber
\end{eqnarray}
which is simply the Taylor expansion of $H(\lambda) = e^{\lambda \eta}
H e^{-\lambda \eta}$ for $\lambda = 1$.

Considering the case where $H = H_0 + \lambda V$, if one can find an
operator $\eta$ such that $[\eta,H_0] = -V$, then using $\lambda \eta$
as a generator leads to
\begin{eqnarray}
\tilde{H} = H_0 + \sum_{n=1}^{\infty} \frac{n\lambda^{n+1}}{(n+1)!}[\eta,
[\eta,\dots [\eta,V] \dots]]. \nonumber
\end{eqnarray}
This operator is a series in powers of $\lambda$, and hence of the
perturbation $V$. While its structure is reminiscent of the results of
high-order perturbation theory, there is an important difference: $\tilde{H}$
acts in the full Hilbert space of $H$, whereas $H_{\rm eff}$ acts in the
ground-state manifold of $H_0$. Depending on the problem, this may or may
not be an advantage. If one is interested in high-energy states which may
be detected in a particular experiment, the canonical-transformation approach
has distinct advantages, because it gives all of the eigenstates up to a
certain order, and not only those which have evolved from the ground-state
manifold of $H_0$ under the perturbation. By contrast, if the goal is to
reduce the Hilbert space to study the low-energy sector using, for example,
exact-diagonalization calculations on finite clusters, then a degenerate
perturbation theory is sufficient.

\subsection{CUT (Continuous Unitary Transformation)}

The canonical transformation introduced in the previous section is by
no means the only possibility for obtaining an effective model by a unitary
transformation. In fact there are many ways to do this, even in low orders
of a perturbative approach, and it is therefore an obvious question to ask
whether an optimal choice of transformation exists.

This question led both Wegner \cite{wegne94} and G{\l}azek and Wilson
\cite{glaze93,glaze94} to introduce independently of each other the concept
of {continuous unitary transformations} ({CUT}s) \cite{knett01b}. In contrast to
the one-step transformation discussed in the last section, here the unitary
transformation is constructed as an infinite product of infinitesimal
transformations. Although measurable (on-shell) quantities, such as energies,
have to be the same independent of which kind of transformation has been
chosen, off-shell quantities such as effective interactions can differ
strongly. This has been demonstrated in a quite impressive manner for the
case of the Fr\"ohlich Hamiltonian, which describes conventional
superconductivity mediated by the electron-phonon interaction \cite{lenz96}.
Here the effective electron-electron interaction at second order in the
electron-phonon coupling shows divergences for the case of a one-step
transformation (previous section), whereas in the continuous treatment
the attractive interaction is smooth.

Another respect in which the approaches differ is that the one-step canonical
transformation can be applied practically only at low orders in the
perturbation. However, there are physical situations where one is interested
in the quantitative determination of an effective Hamiltonian for a given
parameter set in the original model. One example is the case of strongly
frustrated networks of coupled dimers, as in the Shastry-Sutherland model,
where processes relevant to the magnetization of the system appear only
at high orders.

In the method of CUTs, a continuous parameter $l$ is introduced such that
$l = 0$ refers to the initial system $H$ and $l = \infty$ corresponds
to the final effective system, which should correspond to a simplified
physical picture. The transformation can be constructed such that processes
at higher energies are treated before those at lower energies. This
renormalizing property is similar to Wilson's renormalization-group
approach \cite{wilso75}.

Let $U$ be the unitary transformation which diagonalizes the Hamiltonian
$H$ and let $H(l) = U^\dagger (l)HU(l)$. This unitary transformation is then
equivalent to performing an infinite sequence of unitary transformations,
$e^{-\eta (l)dl}$, with the {antihermitian generators}
\begin{equation}
 \eta(l) = - U^\dagger (l) \partial_l U(l).
\end{equation}
Taking the derivative with respect to $l$ results in the ``{flow equation}''
\begin{equation}
 \partial_l H (l) = [\eta (l),H(l)],
 \label{eq:flow}
\end{equation}
which defines the change of the Hamiltonian during the flow.  Note that
Eq.~\ref{eq:flow} represents an infinite hierarchy of coupled differential
equations, because for the general case an increasing number of terms is
generated on the right-hand side at each order. In practice, one has
therefore to perform a {truncation} (below).

The properties of the effective Hamiltonian depend strongly on the choice
of the generator. There are in essence two different modern variants. The
first one uses the generator introduced originally by Wegner \cite{wegne94},
which aims to eliminate interaction matrix elements with the goal of obtaining
an energy-diagonal effective Hamiltonian. This approach has been applied
successfully to a large class of problems, with special attention being
given to determining the ground state of interacting quantum many-body
problems \cite{wegne94,lenz96,kehre06}. The second variant is the
quasiparticle-conserving CUT which, as its name suggests, maps the
Hamiltonian $H_0$ to an effective Hamiltonian conserving the number of
quasiparticles \cite{stein97,mielk98,uhrig98c,knett00a}.
This approach can be used either to study the excitations of an already
known quantum ground state \cite{knett01b,knett00a,knett00b,heidb02,breni03,
schmi05}, one application for which has been to bound states, or, in analogy
to the previous case, to derive effective low-energy models
\cite{reisc03a,dusue04a,dusue04,kriel05,schmi08}.

Returning to the point mentioned above, in all practical calculations it is
necessary to truncate the flow equation (\ref{eq:flow}). For this there are
two options: i) cutting the hierarchy at one level and solving the equations
numerically (self-similar CUTs) or ii) using a series-expansion ansatz for
$\eta$ and $H$, and solving the flow equations perturbatively to high order
({perturbative CUTs}).

Here we focus only on presenting one illustrative example, for which
we choose the perturbative version of quasiparticle-conserving CUTs
\cite{stein97,knett00a,knett03a,knett04}.
If the problem at hand meets the two conditions:
\begin{enumerate}
 \item the unperturbed part has an equidistant spectrum bounded from below,
 \item there is an integer number $N > 0$ such that the perturbing part can
be separated as $\sum_{n = -N}^{N} T_n$, where $T_n$ increments (or
if $n < 0$ decrements) the number of particles by $n$,
\end{enumerate}
then the CUT in its quasiparticle-conserving form can be solved to high
order in the perturbation and the effective Hamiltonian is given by the
general form \cite{knett00a}
\begin{equation}
  \label{H_eff}
  H_{\rm eff}(x) = Q + \sum_{k=1}^{\infty}x^{k}
\sum_{|\underline{m}| = k \atop M(\underline{m})=0}  C(\underline{m})
T(\underline{m}),
\end{equation}
where $Q$ is the unperturbed part of the Hamiltonian, $x$ an expansion
parameter, $\underline{m} = (m_1,m_2,\ldots,m_k)$, and $M(\underline{m}) =
\sum_{i=1}^k m_i = 0$ reflects the conservation of the number of particles.
The action of $H_{\rm eff}$ can be viewed as a weighted sum of
particle-number-conserving virtual excitation processes, each of which
is encoded in a monomial $T(\underline{m}) = T_{m_1} T_{m_2}  \ldots T_{m_k}$.
The coefficients $C(\underline{m})$ are rational numbers which can be
calculated (to high order in the perturbation) exactly as the ratio of two
integers. It should be emphasized that the effective Hamiltonian $H_{\rm
eff}$, which has known coefficients $C(\underline{m})$, can be used
straightforwardly in all perturbative problems that meet the above
conditions.

We illustrate the method first for an unfrustrated {spin ladder} in a magnetic
field, including from Eq.~(\ref{eq:ladder}) only the magnetic exchange
$J_{\perp}$ on the rungs, the magnetic field $B$, and the unfrustrated
exchange along the legs of the ladder, setting $J_{11} = J_{22} \equiv
J_{\parallel}$. Rewriting the Hamiltonian in terms of rung triplet
operators $t_\alpha$, with $\alpha = \{ x,y,z\}$ \cite{sachd90}, we obtain
$$
H = J_\perp Q + \frac{J_\parallel}{2}\left[ T_0 + T_{+2} + T_{-2} \right] +
H_B,
$$
where
\begin{eqnarray*}
 Q & = & \sum_{i,\alpha} t^{\dagger}_{\alpha,i} t_{\alpha,i},\\
 T_0 & = & \sum_{i,\alpha} t^{\dagger}_{\alpha,i} t_{\alpha,i+1} +
\sum_{i,\alpha,\beta} \left[  t^{\dagger}_{\alpha,i}  t^{\dagger}_{\beta,i+1}
t_{\beta,i} t_{\alpha,i+1} - t^{\dagger}_{\alpha,i}  t^{\dagger}_{\alpha,i+1}
t_{\beta,i} t_{\beta,i+1} \right], \\
T_{+2} &=& \sum_{i,\alpha} t^{\dagger}_{\alpha,i} t^{\dagger}_{\alpha,i+1},\\
T_{-2} &=& \sum_{i,\alpha} t_{\alpha,i} t_{\alpha,i+1} = T_{+2}^\dagger.
\end{eqnarray*}
The operator $Q$ counts the total number of triplet excitations, while the
operators $T_n$ change the triplet number by $n$, and $H_B$ denotes the
magnetic-field term.

In the following we consider the limit of weakly coupled rung dimers, i.e.
we set $J_\perp = 1$ and consider $J_\parallel/J_\perp \equiv x$ as the small
expansion parameter. The effective Hamiltonian $H_{\rm eff}$ obtained by a
{quasiparticle-conserving CUT} has the property $[H_{\rm eff},Q] = 0$, meaning
that the total number of {triplons} (dressed triplets which are the elementary
excitations of the spin ladder) is a conserved quantity. The effective
Hamiltonian at second order is
$$
H_{\rm eff}^{(2)} = Q + x T_0 + \frac{x^2}{4} \left[T_{+2} T_{-2}
 - T_{-2} T_{2} \right] + H_B.
$$
The total spin $S^z_{\rm tot}$ is a conserved quantity. The magnetic-field
term is therefore not changed under the unitary transformation, and the
low-energy physics is influenced solely by the local singlet $|s \rangle$
and the triplet $|t^1 \rangle$ polarized parallel to the magnetic field (as
discussed in Sec.~2). Identifying $|s\rangle$ with an empty site and
$|t^1 \rangle$ with the presence of a hard-core boson, one may deduce the
effective Hamiltonian in this basis by calculating matrix elements on a
finite cluster \cite{knett03a,knett04}. The result is
\begin{eqnarray}
H_{\rm eff}^{hb} & = & -t_n \sum_{i,j\in\{1,2\}}\left( b^\dagger_i b_{i+j}
+ {\rm h.c.}\right) -\mu \sum_i n_i \nonumber\\
&& - t'_1 \sum_{i} \left( b^\dagger_{i-1} b_{i+1} + {\rm h.c.}
\right) n_i + v_j\sum_{i,j\in\{1,2\}} n_i n_{i+j},
\end{eqnarray}
with $t_1 = x/2$, $t_2 = x^2/4$, $t'_1 = -x^2/4$, $v_1 = x/2 - 3x^2/8$,
$v_2 = 0$, and $\mu = B - 1 + 3x^2/4$.

We emphasize that a calculation such as this, based on coupled dimers in
an external magnetic field, can be performed for any frustrated lattice and
to high orders in the perturbation. By using appropriate extrapolations, a
quantitative low-energy effective Hamiltonian may be derived, which is usually
applicable over a large part of the parameter space. The high-order expansion
(including extrapolation of the bare series) becomes problematic only when
the correlation length of the system exceeds the spatial range covered by
the maximum order treated, for example close to a quantum phase transition.
The method will break down if the ground state for the parameters of interest
is not unitarily connected to the ground state about which one is expanding.

\begin{figure}
   \begin{center}
   \includegraphics[width=0.6\columnwidth]{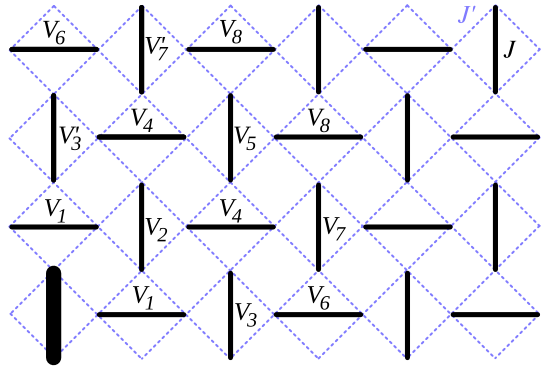}
   \end{center}
    \caption{Shastry-Sutherland lattice and definition of the two-body
interactions. $V_n$ is the coefficient of the two-body interactions
between the reference dimer (thick bond) and the corresponding dimer.
Figure courtesy of Ref.~\cite{dorier08}.}
    \label{fig:lattice}
\end{figure}

As a second example illustrating the importance of a quantitative effective
model, meaning one obtained with high-order accuracy, we discuss the 2D
spin-1/2 Heisenberg system known as the {Shastry-Sutherland model}
\cite{shastry82} in a magnetic field,
\begin{equation}
H=J'\sum_{\langle i,j \rangle}\, S_{i}\cdot \, S_{j}+J\sum_{\ll i,j\gg}\,
S_{i}\cdot \, S_{j}-B\sum_{i}S_i^z.
\end{equation}
The bonds denoted $\ll i,j\gg$ represent an array of orthogonal dimers, while
the bonds $\langle i,j \rangle$, which are inter-dimer couplings, form a
square lattice (Fig.~\ref{fig:lattice}). This theoretical model is believed
to be realized experimentally in the layered copper oxide material
SrCu$_2$(BO$_3$)$_2$, where the coupling ratio is $J'/J\approx 0.63$.
In the theoretical model, for $J'/J$ smaller than a critical ratio of
order 0.7, the ground state of the model is given exactly by the product
of dimer singlets, and the magnetization process of the system can be
described in terms of hard-core bosons which, as discussed for the spin
ladder above, represent polarized $|t^1\rangle$ triplons on the dimers,
interacting and moving on an effective square lattice
\cite{momoi00,miyahara03R}.

A consequence of the strong frustration is the appearance of several
{magnetization plateaus} which correspond to Mott-insulating phases of the
hard-core bosons \cite{muell00,onizuka00,kodama02}, where the translational symmetry
of the system is broken and triplon excitations are frozen in the ground state
as in a charge-ordered state (discussed in the chapter by Takigawa and
Mila). Theoretically, all approaches to the basic theoretical model agree on
the presence of magnetization plateaus at 1/3 and 1/2 of the saturation
value \cite{momoi00,misguich01,miyahara00,miyahara03R,miyahara03}, in
agreement with experiments \cite{onizuka00,sebastian07}. However, the
structure below 1/3 magnetization is rather controversial. On the
experimental side, the original pulsed-field data show only two anomalies
which were interpreted as plateaus at 1/8 and 1/4 \cite{onizuka00}, but the
presence of additional phase transitions, and of a broken translational
symmetry above the 1/8-plateau have been established by recent torque and
NMR measurements performed up to 31~T \cite{levy08,Takigawa07}. The
possibility of additional plateaus has been pointed out by Sebastian
{\it et al.}~\cite{sebastian07}, who interpreted their high-field torque
measurements as evidence for plateaus at $1/q$, with $2\le q\le 9$, and
at $2/9$. On the theoretical side, the situation is also not settled.
The finite clusters available for exact-diagonalization studies are not
large enough to allow reliable predictions for high-commensurability plateaus,
and the accuracy of the Chern-Simons mean-field approach initiated by Misguich
{\it et al.}~\cite{misguich01}, and employed recently by Sebastian {\it et
al.}~\cite{sebastian07} to explain their apparent additional plateaus, is
not easy to assess. The essential difficulty lies in the fact that, because
plateaus are a consequence of repulsive interactions between triplons, an
accurate determination of the low-density, high-commensurability plateaus
requires a precise knowledge of the long-range part of the interaction.

Such a precise analysis was conducted recently using perturbative CUTs
\cite{dorier08}. The processes relevant for the physics in a finite
magnetic field are those with maximum total spin and total $S_z$. Thus
the general form of the effective Hamiltonian obtained by the perturbative
CUT takes the form
\begin{equation}
H_{\rm eff} = \sum_{n=2,4,6 \dots}\, \sum_{r_1,\dots,r_n} C_{r_1,\dots,r_n}
b_{r_1}^\dag \dots b_{r_{n/2}}^\dag b_{r_{n/2+1}} \dots b_{r_n},
\end{equation}
where $r_i$ labels the sites of the square lattice formed by the $J$ bonds,
while the hard-core boson operator $b_r^\dag$ creates a polarized $|t^1
\rangle$ triplon at site $r$. The coefficients $C_{r_1,r_2,\dots,r_n}$ are
then obtained as high-order series in $J'/J$, computed up to order 15 for
the two-body interactions $V_n$ to be discussed below.

\begin{figure}
   \begin{center}
     \includegraphics[width=0.8\columnwidth]{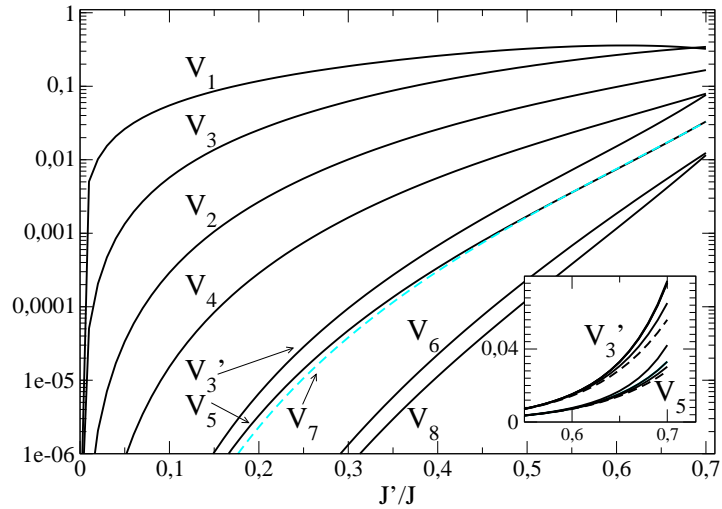}
   \end{center}
   \caption{Coefficients of the extrapolated two-body interactions $V_n$
as function of $J'/J$. Inset: different extrapolations (solid lines) shown
together with the bare series (dashed lines) for $V'_3$ and $V_5$. Figure
courtesy of Ref.~\cite{dorier08}.}
    \label{fig:hamiltonien_twobody_Jp}
\end{figure}

It is found that the magnitudes of all the interaction terms decrease when
the separation of the sites is increased. In addition, the physics at low
density is dominated by the two-body density-density interactions, while the
standard two-site hopping is, as expected, strongly suppressed due to the
frustration \cite{miyahara99}. The evolution with $J'/J$ of the most relevant
two-body interactions, defined in Fig.~\ref{fig:lattice}, is shown in
Fig.~\ref{fig:hamiltonien_twobody_Jp}. At small $J'/J$, interactions beyond
$V_4$ are small and may be neglected, but for larger coupling ratios the
higher-order terms $V'_3$, $V_5$, and $V_7$ (appearing at order 6) become
important and contribute to the formation of low-density plateaus.

In general, the effective Hamiltonian $H_{\rm eff}$ is by no means simpler
than the original one, but it becomes so in the limit of low density and
moderate $J'/J$. Indeed, in this limit the kinetic terms are very small,
and they can be considered as a perturbation of the interaction part,
which is diagonal in the local Fock basis, $|n_{r_1}, n_{r_2}, \dots \rangle$.
It is thus appropriate to use a {Hartree approximation} in which the variational
ground state is a product of local boson wave-functions, because this
approximation becomes exact in the limit of vanishing kinetic energy. This
approach has the further advantage that it can be used to compare and treat
rather large unit cells, which is important in the SrCu$_2$(BO$_3$)$_2$
problem, where ``solid'' phases with a complicated structure can be found at low magnetizations.

\begin{figure}
   \begin{center}
   \includegraphics[width=0.8\columnwidth]{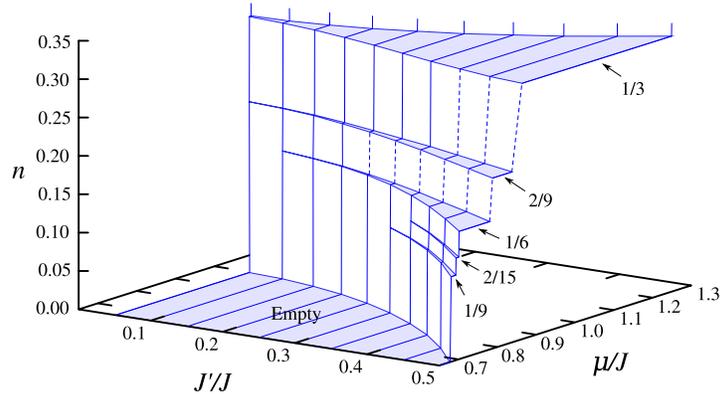}
   \end{center}
    \caption{Magnetization plateaus as a function of $\mu$ and $J'/J$. The
boson density $n$ is equal to the magnetization in units of the saturation
value, and the chemical potential $\mu$ is equal to the magnetic field $B$.
Solid lines denote results that are fully converged with respect to the
terms retained in the Hamiltonian. Well-converged results are then connected
by dashed lines. Figure courtesy of Ref.~\cite{dorier08}.}
    \label{fig:phasediagram}
\end{figure}

The resulting phase diagram in the Hartree approximation is shown in
Fig.~\ref{fig:phasediagram}. The phase diagram is dominated by a series
of plateaus, which appear at 1/3 and 1/2 (not shown) even at very small
$J'/J$, and at 2/9, 1/6, 1/9, and 2/15 as $J'/J$ is increased. A plateau
structure of this kind is to be expected when the kinetic terms, as here,
are very small, because if these terms were completely absent, the
magnetization curve would consist simply of a sequence of plateaus.
At $J'/J = 0.5$, the 1/6-plateau is by far the most prominent stucture
below 1/3.

All of the plateaus found at low densities are actually stabilized by two-body
repulsive interactions $V_n$ appearing at high orders in $J'/J$. It is
therefore crucial to obtain the effective Hamiltonian very accurately,
because the most significant features in this density regime are found to
result from the competition between small interactions.

Finally, it is worth adding that the method of CUTs is also able to treat
{observables} \cite{kehre06,knett03a,dorier08}. To this end, an observable $O$ must be transformed by the
same CUT as the Hamiltonian,
\begin{equation}
 \partial_l O (l) = [\eta (l),O(l)],
 \label{eq:ons}
\end{equation}
yielding effective observables $O_{\rm eff} = O(l=\infty)$. Here we
mention only one possible application which can be very useful for the
physics of frustrated quantum magnets: a typical situation in quantum
magnetism is that, in addition to the dominant nearest-neighbor Heisenberg
exchange interactions, there are small coupling terms $H_{\rm add}$, such
as {Dzyloshinskii-Moriya interactions}, which can have a profound influence
on the physics of the system. Formally
$$
 H_{\rm tot} = H + H_{\rm add},
$$
and it is both elegant and efficient to perform first a CUT on $H$,
which contains the dominant couplings, and then to treat $H_{\rm add}$
as an observable
$$
 H_{\rm tot}^{\rm eff} = H^{\rm eff} + U^\dagger H_{\rm add} U.
$$
Here $H^{\rm eff}$ conserves the number of quasiparticles, whereas
the transformed observable $U^\dagger H_{\rm add} U$ does not. Thus in the
second step, performed after the first CUT, either ordinary perturbation
theory or a second CUT can be applied to treat the term $U^\dagger H_{\rm
add} U$, which contains the small couplings. In cases where $H_{\rm add}$
mixes low- and high-energy states, it is essential to retain access to the
full Hilbert space of the problem.

\subsection{CORE ({Contractor Renormalization})}

One further method which we highlight here for the derivation of
low-energy effective models is the {CORE} technique, invented by Morningstar
and Weinstein \cite{morni94}. The underlying idea is to derive effective
Hamiltonians for a truncated local basis in such a manner that the
low-energy spectrum of the model under study is reproduced exactly.
As for CUTs, the CORE approach can be used either in an analytically
oriented form \cite{weins01,altma02,berg03,budni04,siu07} or as a numerical
technique \cite{pieka97,cappo02,cappo04,abend07}. For recent reviews on CORE
we refer the reader to Refs.~\cite{auerb06,cappo06}.

The essential steps of the CORE algorithm are the following.

\noindent
1) The system is divided into local subunits. One subunit is diagonalized,
keeping $M$ suitable low-energy states.\\
2) The full Hamiltonian is diagonalized on a connected graph consisting of
$N$ subunits. The low-energy eigenenergies $\epsilon_n$ and eigenstates
$|n\rangle$ are calculated.\\
3) A basis of dimension $M^N$ is obtained by projecting the eigenstates
onto the tensor product space of the retained states.\\
4) The effective Hamiltonian is constructed according to
$$
 H^{\rm eff}_N = \sum_{n=1}^{M^N} \epsilon_n |\psi_n\rangle\langle\psi_n|.
$$
5) The connected range-$N$ interactions are determined by subtracting the
contributions of all connected subclusters.\\
Finally, the effective Hamiltionian is deduced by a cluster expansion as
$$
 H^{\rm eff}_{\rm CORE} = \sum_i H_i + \sum_{\langle i,j\rangle} H_{ij} +
\ldots
$$
Note that $H^{\rm eff}_{\rm CORE}$ reproduces exactly the low-energy physics
if one considers all of the terms on the right-hand side.

In practice, it is necessary to perform a {truncation}. The convergence of
the algorithm therefore depends both on the range of the operators taken
in the cluster expansion and on the number and type of low-energy states
retained for one subunit. Hence the successful application of the CORE
technique does require some physical insight concerning the problem at
hand. However, once the relevant degrees of freedom have been selected,
CORE represents a non-perturbative method for deriving effective low-energy
Hamiltonians.

An important feature of the CORE algorithm is that it does not rely on the
system being in a certain physical phase (to be contrasted with the example
of quasiparticle-conserving CUTs discussed in the preceding section) and
therefore does not break down even if a quantum phase transition takes place
in the parameter space of the original model. As an example, we present here
some illustrative CORE results for the magnetization curves of the 2D
spin-1/2 Heisenberg model on the {Shastry-Sutherland lattice} discussed
in the previous section \cite{capponi08}. The CORE technique is used to
derive an effective Hamiltonian which is then treated by exact diagonalization
(ED). The effective model obtained by CORE is found to agree very well with
that obtained by perturbative CUTs. It is therefore expected that differences
between the two approaches arise primarily from the method used to treat the
effective model, which in the examples shown is either a classical limit
\cite{dorier08} or ED \cite{capponi08}.

\begin{figure}
   \begin{center}
     \includegraphics[width=0.8\columnwidth]{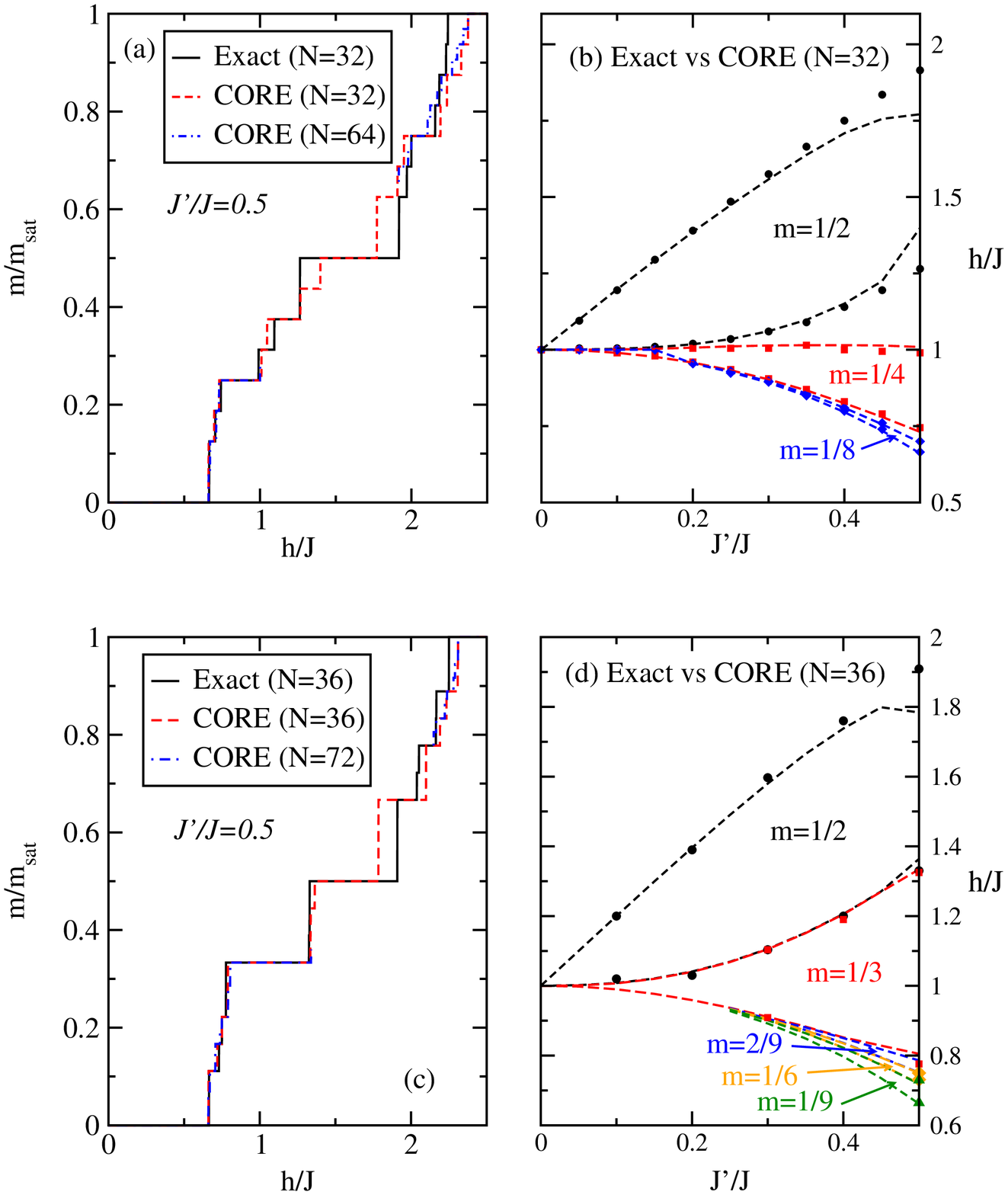}
   \end{center}
   \caption{(a) Magnetization curves obtained by ED and CORE calculations
on an $N = 32$ lattice and from CORE results with $N = 64$ for $J'/J = 0.5$.
(b) Phase diagram for $N = 32$ as a function of $J'/J$ and magnetic field
$h/J$: CORE results (lines) agree with ED (symbols) for locations of the
$m = 1/2$, $1/4$, and $1/8$ plateaus which are allowed on this cluster.
(c,d) Same as (a,b) for $N = 36$ and $N = 72$. On these clusters, the
$m = 1/2$, $1/3$, $2/9$, $1/6$, and $1/9$ plateaus are allowed. Figure
courtesy of Ref.~\cite{capponi08}.}
    \label{fig:capponi_core}
\end{figure}

Results obtained from CORE+ED are shown in Fig.~\ref{fig:capponi_core}.
As in the treatment by a perturbative CUT, a rich plateau structure is
found below $m = 1/3$. This again highlights the utility of an approach
to the physics by first deriving a quantitative effective model which is
then treated by simpler techniques. One obvious advantage of ED compared
to the classical treatment is that it takes quantum fluctuations fully
into account. It is therefore striking that the magnetization curves in
both approaches are dominated by the presence of plateaus. By contrast,
one drawback in using ED as a solver for the effective model (and one
advantage of the classical solver) is the restriction on cluster sizes
and shapes. Because the physics at low densities involves solid phases
with rather large unit cells, it is a challenge within the ED approach
to stabilize and to compare different solids, such as the $2/15$ phase
found in Ref.~\cite{dorier08}. Despite the differing aspects, both
positive and negative, of these approaches to this problem, it should
be emphasized that the advanced techniques used to derive effective
Hamiltonians have proven to be crucial in the discovery and resolution
of the complicated {magnetization processes} of strongly frustrated quantum
magnets.

\section{Conclusions}

In this chapter we have summarized a variety of tools which are used in
the field of highly frustrated magnetism to derive effective {low-energy
Hamiltonians}. We have aimed to capture the essential technical aspects
of these different approaches and to provide examples of them for a number
of physical applications. While it is not possible in a chapter of this
type to cover all such methods in full detail, where relvant we have
referred the interested reader to the more specialized literature.

We have shown that strong-coupling expansions and the derivation of
effective models are at the same time standard techniques used by
theoretical physicists for a broad range of physical questions and also
a very active area of current research on strongly correlated quantum
systems. For the first type of application, it is often the goal to
identify the relevant low-energy degrees of freedom and to use an
effective model to leading order in perturbation theory to understand
qualitatively the phase diagram of a given model. By contrast, the aim
of the current developments is to obtain a quantitative derivation of
effective models and a complete understanding of the breakdown of such
derivations.

\printindex


\begin{thebibliography}{99.}

\bibitem{landau} See for example L. Landau and E. Lifshitz, {\sl Quantum 
mechanics}, (Pergamon, 1991).

\bibitem{fulde} See for example P. Fulde, {\sl Electron Correlations in 
Molecules and Solids}, (Springer, 1991).

\bibitem{kato} T. Kato, Prog. Theor. Phys. \textbf{4}, 514 (1949).

\bibitem{takahashi} M. Takahashi, J. Phys. C \textbf{10}, 1289 (1977).

\bibitem{anderson} P. W. Anderson, Phys. Rev. \textbf{115}, 2 (1959).

\bibitem{totsuka} K. Totsuka, Phys. Rev. B \textbf{57}, 3454 (1998).

\bibitem{mila} F. Mila, Eur. Phys. J. B \textbf{6}, 201 (1998).

\bibitem{bethe} H. A. Bethe, Z. Phys. \textbf{71}, 205 (1931).

\bibitem{luther} A. Luther and I. Peschel, Phys. Rev. B \textbf{9}, 2911 
(1974).

\bibitem{haldane} F. D. M. Haldane, Phys. Rev. Lett. \textbf{45}, 1358 (1980).

\bibitem{mattis} See for example D. C. Mattis, {\sl The theory of magnetism 
made simple}, (World Scientific, 2006).

\bibitem{fouet} J.-B. Fouet, F. Mila, D. Clarke, H. Youk, O. Tchernyshyov, 
P. Fendley, and R. M. Noack, Phys. Rev. B \textbf{73}, 214405 (2006).

\bibitem{chandra} R. Moessner, S. L. Sondhi, and P. Chandra, Phys. Rev. Lett. 
\textbf{84}, 4457 (2000).

\bibitem{misguich} G. Misguich and F. Mila, Phys. Rev. B \textbf{77}, 134421 
(2008).

\bibitem{cabra}  D. C. Cabra, M. D. Grynberg, P. C. Holdsworth, A. Honecker, 
P. Pujol, J. Richter, D. Schmalfuss, and J. Schulenburg, Phys. Rev. B 
\textbf{71}, 144420 (2005).

\bibitem{bergman} D. L. Bergman, R. Shindou, G. A. Fiete, and L. Balents,
Phys. Rev. B \textbf{75}, 094403 (2007).

\bibitem{rousochatzakis} I. Rousochatzakis, A. M. L\"auchli, and F. Mila,
Phys. Rev. B \textbf{77}, 094420 (2008).

\bibitem{schulz} H. J. Schulz in {\sl Proceedings of the XXXIst Rencontres de 
Moriond}, edited by T. Martin, G. Montambaux, and J. Trin Thanh Vin (Editions 
Fronti\`eres, Gif-sur-Yvette, 1996).

\bibitem{subrahmanyam} V. Subrahmanyam, Phys. Rev. B \textbf{52}, 1133 (1995).

\bibitem{mila2}  F. Mila, Phys. Rev. Lett. \textbf{81}, 2356 (1998).

\bibitem{ferrero} For a detailed discussion of this point, see M. Ferrero, 
F. Becca, and F. Mila, Phys. Rev. B \textbf{68}, 214431 (2003).

\bibitem{lecheminant} P. Lecheminant, B. Bernu, C. Lhuillier, L. Pierre, 
and P. Sindzingre, Phys. Rev. B \textbf{56}, 2521 (1997).

\bibitem{honecker} B. Damski, H.-U. Everts, A. Honecker, H. Fehrmann, 
L. Santos, and M. Lewenstein, Phys. Rev. Lett. \textbf{95}, 060403 (2005).

\bibitem{tsunetsugu}  H. Tsunetsugu, Phys. Rev. B  \textbf{65}, 024415 (2002).

\bibitem{kotov} V. N. Kotov, M. E. Zhitomirsky, M. Elhajal, and F. Mila,
Phys. Rev. B \textbf{70}, 214401 (2004);  V. N. Kotov, M. Elhajal, 
M. E. Zhitomirsky, and F. Mila, Phys. Rev. B \textbf{72}, 014421 (2005).

\bibitem{schmi00} H. J. Schmidt and Y. Kuramoto, Physica C \textbf{167}, 
263 (1990).

\bibitem{takah77} M. Takahashi, J. Phys. C: Solid State Phys. \textbf{10}, 
1289 (1977).

\bibitem{macdo90} A. H. MacDonald, S. M. Girvin, and D. Yoshioka, Phys. 
Rev. B \textbf{41}, 2565 (1990).

\bibitem{muell02} E. M\"uller-Hartmann and A. Reischl, Eur. Phys. J. B 
\textbf{28}, 173 (2002).

%CUT

\bibitem{wegne94} F. J. Wegner, Ann. Physik \textbf{3}, 77 (1994).

\bibitem{glaze93} S. D. G{\l}azek and K. G. Wilson, Phys. Rev. D 
\textbf{48}, 5863 (1993).

\bibitem{glaze94} S. D. G{\l}azek and K. G. Wilson, Phys. Rev. D 
\textbf{49}, 4214 (1994).

\bibitem{knett01b} C. Knetter, K. P. Schmidt, M. Gr\"uninger, and 
G. S. Uhrig, Phys. Rev. Lett. \textbf{87}, 167204 (2001).

\bibitem{lenz96} P. Lenz and F. Wegner, Nucl. Phys B \textbf{482}, 693 (1996).

\bibitem{wilso75} K. G. Wilson, Rev. Mod. Phys. \textbf{47}, 773 (1975).

\bibitem{kehre06} S. Kehrein, {\sl Flow-equation approach to many-particle 
systems}, Springer Tracts in Modern Physics {217} (2006), and references 
therein.

%Quasi-particle CUT

\bibitem{stein97} J. Stein, J. Stat. Phys. \textbf{88}, 487 (1997).

\bibitem{mielk98} A. Mielke, Eur. Phys. J. B \textbf{5},  605  (1998).

\bibitem{uhrig98c} G. S. Uhrig and B. Normand, Phys. Rev. B \textbf{58}, 
R14705 (1998).

\bibitem{knett00a} C. Knetter and G. S. Uhrig, Eur. Phys. J. B \textbf{13}, 
209 (2000).

\bibitem{knett00b} C. Knetter, A. B\"uhler, E. M\"uller-Hartmann, and G. S. 
Uhrig, Phys. Rev. Lett. \textbf{85}, 3958 (2000).

\bibitem{heidb02} C. P. Heidbrink and G. S. Uhrig, Eur. Phys. J. B 
\textbf{30}, 443 (2002).

\bibitem{breni03} W. Brenig, Phys. Rev. B \textbf{67}, 064402 (2003).

\bibitem{schmi05} K. P. Schmidt and G. S. Uhrig, Mod. Phys. Lett. B 
\textbf{19}, 1179 (2005).

\bibitem{reisc03a} A. Reischl, E. M\"uller-Hartmann, and G. S. Uhrig, 
Phys. Rev. B \textbf{70}, 245124 (2004).

\bibitem{dusue04a} S. Dusuel and G. S. Uhrig,  J. Phys. A: Math. and Gen. 
\textbf{37}, 9275 (2004).

\bibitem{dusue04} S. Dusuel and J. Vidal, Phys. Rev. Lett. \textbf{93}, 
237204 (2004).

\bibitem{kriel05} J. N. Kriel, A. Y. Morozov, and F. G. Scholtz, J. Phys. 
A: Math. and Gen. \textbf{38}, 205 (2005).

\bibitem{schmi08} K. P. Schmidt, S. Dusuel, and J. Vidal, Phys. Rev. Lett. 
{\bf 100}, 057208 (2008).

\bibitem{knett03a} C. Knetter, K. P. Schmidt, and G. S. Uhrig, J. Phys.: 
Condens. Matter \textbf{36}, 7889 (2003).

\bibitem{knett04} C. Knetter, K. P. Schmidt, and G. S. Uhrig, Eur. Phys. 
J. B \textbf{36}, 525 (2004).

\bibitem{sachd90}
S. Sachdev and R. N. Bhatt, Phys. Rev. B \textbf{41}, 9323 (1990).

\bibitem{shastry82} B. S. Shastry and B. Sutherland, Physica B {\bf 108B}, 
1069 (1981).

\bibitem{muell00} E. M\"uller-Hartmann, R. R. P. Singh, C. Knetter, and G. S. Uhrig, Phys. Rev. Lett. \textbf{81}, 1808 (2000).

\bibitem{onizuka00} K. Onizuka {\it et al.}, J. Phys. Soc. Jpn. {\bf 69}, 
1016 (2000).

\bibitem{kodama02} K. Kodama {\it et al.}, Science {\bf 298}, 395 (2002).

\bibitem{momoi00} T. Momoi and K. Totsuka, Phys. Rev. B {\bf 62}, 15067 (2000).
\bibitem{miyahara03R}
S. Miyahara and K. Ueda, J. Phys.: Condens. Matter {\bf 15}, R327 (2003).

\bibitem{miyahara00}
S. Miyahara and K. Ueda, Phys. Rev. B {\bf 61}, 3417 (2000).

\bibitem{misguich01} G. Misguich, T. Jolicoeur, and S. M. Girvin, Phys. 
Rev. Lett. {\bf 87}, 097203 (2001).

\bibitem{miyahara03}
S. Miyahara, F. Becca and F. Mila, Phys. Rev. B {\bf 68}, 024401 (2003).

\bibitem{sebastian07} S. E. Sebastian, N. Harrison, P. Sengupta, C. D. 
Batista, S. Francoual, E. Palm, T. Murphy, H. A. Dabkowska, and B. D. Gaulin, 
Proc. Nat. Acad. Sci. \textbf{105}, 20157 (2008).

\bibitem{levy08} F. Levy, I. Sheikin, C. Berthier, M. Horvati\'{c}, 
M. Takigawa, H. Kageyama, T. Waki, and Y. Ueda, Europhys. Lett. \textbf{81}, 
67004 (2008).

\bibitem{Takigawa07}  M. Takigawa, S. Matsubara, M. Horvati\'{c}, C. Berthier, 
H. Kageyama, and Y. Ueda, Phys. Rev. Lett. \textbf{101}, 037202 (2008).

\bibitem{dorier08} J. Dorier, K. P. Schmidt, and F. Mila, Phys. Rev. Lett. 
{\bf 101}, 250402 (2008).

\bibitem{miyahara99}
S. Miyahara and K. Ueda, Phys. Rev. Lett. {\bf 82}, 3701 (1999).

\bibitem{capponi08}
A. Abendschein and S. Capponi, Phys. Rev. Lett. {\bf 101}, 227201 (2008).

%CORE

\bibitem{morni94} C. J. Morningstar and M. Weinstein, Phys. Rev. Lett. 
\textbf{73}, 1873 (1994); Phys. Rev. D {\bf 54}, 4131 (1996).

\bibitem{weins01} M. Weinstein, Phys. Rev. B {\bf 63}, 174421 (2001).

\bibitem{altma02}
E. Altman and A. Auerbach, Phys. Rev. B \textbf{65}, 104508 (2002).

\bibitem{berg03}
E. Berg, E. Altmann, and A. Auerbach, Phys. Rev. Lett. {\bf 90}, 147204 (2003).

\bibitem{budni04}
R. Budnik and A. Auerbach, Phys. Rev. Lett. \textbf{93}, 187205 (2004).

\bibitem{siu07}
M. S. Siu and M. Weinstein, Phys. Rev. B \textbf{75}, 184403 (2007).

\bibitem{pieka97} J. Piekarewicz and J. R. Shepard, Phys. Rev. B \textbf{56}, 
5366 (1997);  Phys. Rev. B \textbf{57}, 10260 (1998).

\bibitem{cappo02}
S. Capponi and D. Poilblanc, Phys. Rev. B \textbf{66}, 180503(R) (2002).

\bibitem{cappo04} S. Capponi, A. L\"auchli, and M. Mambrini, Phys. Rev. B 
{\bf 70}, 104424 (2004).

\bibitem{abend07}
A. Abendschein and S. Capponi, Phys. Rev. B \textbf{76}, 064413 (2007).

\bibitem{auerb06} A. Auerbach, AIP Conf. Proc. \textbf{816} 1 (2006).

\bibitem{cappo06} S. Capponi AIP Conf. Proc. \textbf{816} 16 (2006).

\end{thebibliography}
\end{document}